\documentclass[aps,showpacs,superscriptaddress]{revtex4-1}  

\usepackage{url}
\usepackage[utf8]{inputenc}
\usepackage[T1]{fontenc}
\usepackage[english]{babel}

\usepackage{graphicx} 

\usepackage{siunitx} 
\newcommand{\ltotperc}{\eta}
\newcommand{\isoidx}{\beta_0^{2,0}}

\begin{document}

\title{Structural similarity between dry and wet sphere packings}

\author{Simon Weis$^1$, Gerd E.\ Schr\"oder-Turk$^{1,2,3}$, Matthias Schr\"oter$^{4,5}$}
\address{$^1$ Institut für Theoretische Physik I, Friedrich-Alexander-Universität Erlangen-N\"urnberg, 91058 Erlangen, Germany} 
\address{$^2$ Maths \& Stats, School of Engineering and Information Technology, Murdoch University, Perth, Australia}
\address{$^3$ Applied Maths, Research School of Physical Sciences \& Engin., Australian National Univ., Canberra, Australia}
\address{$^4$Institute for Multiscale Simulation, Friedrich-Alexander-Universität Erlangen-N\"urnberg, 91052 Erlangen, Germany}
\address{$^5$Max Planck Institute for Dyamics and Self-Organization, 37077 G\"ottingen, Germany}
\footnote{matthias.schroeter@ds.mpg.de}

\date{\today}

\begin{abstract}
  The mechanical properties of granular materials change significantly in the presence of a wetting liquid which creates capillary bridges between the particles.
  Here we demonstrate, using X-ray tomographies of dry and wet sphere packings,
  that this change in mechanical properties is not accompanied by structural differences between the packings.
  We characterize the latter by the average numbers of contacts of each sphere $\langle Z\rangle$ 
  and the shape isotropy  $\langle \beta_0^{2,0} \rangle$  of the Voronoi cells of the particles.
  Additionally, we show that the number of liquid bridges per sphere $\langle B\rangle$ is approximately 
  equal to $\langle Z\rangle + 2$, independent of the volume fraction of the packing. 
  These findings will be helpful in guiding the development of both particle-based models and continuum mechanical 
  descriptions of wet granular matter.
\end{abstract}

\maketitle

\section{Introduction}

\begin{figure}[b]
    \centering
    \includegraphics[width=\columnwidth]{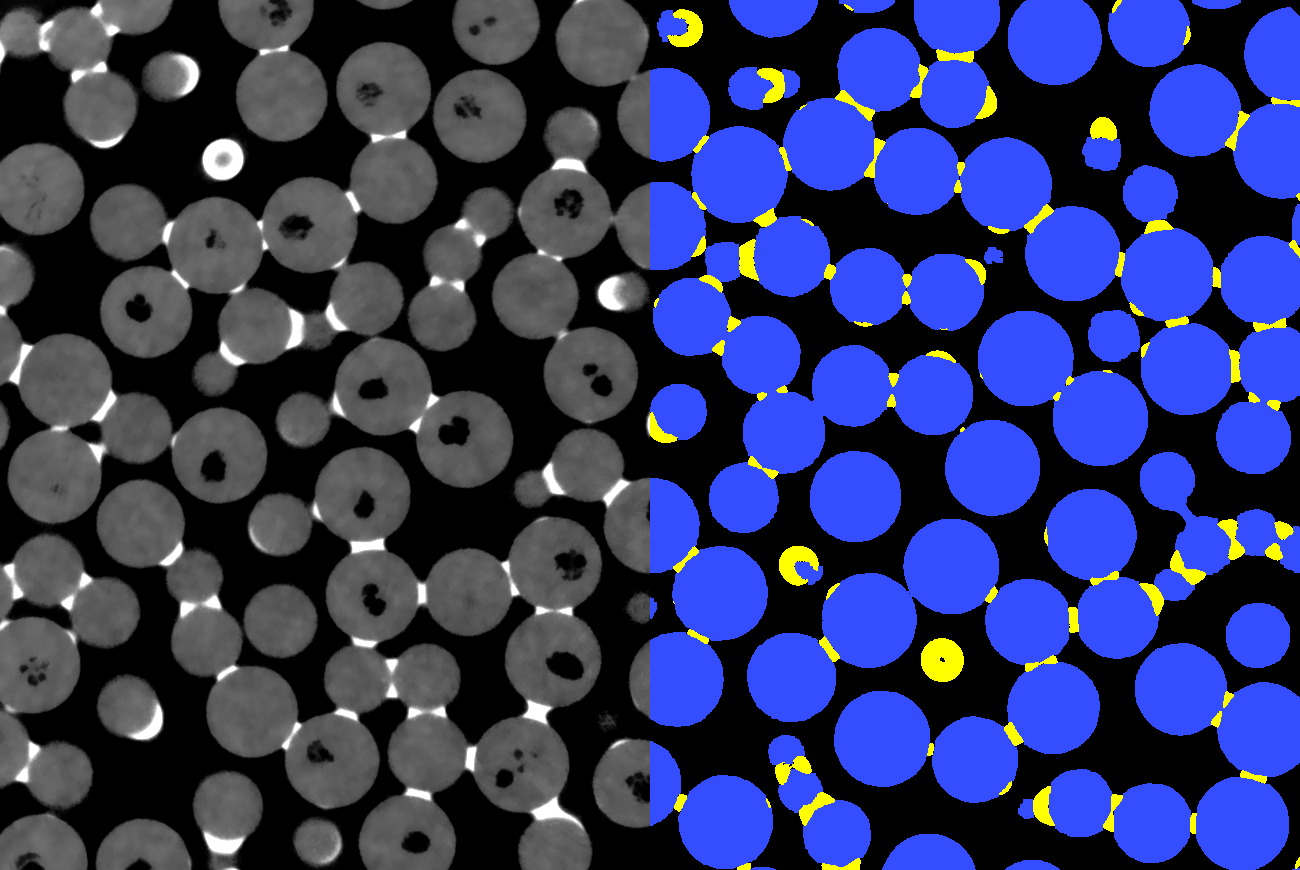}
    
    \caption{Horizontal slice through a tomogram of a wet packing with a liquid content of 3.1\%.
      \textit{Left}: Raw data after applying a bilateral filter for noise reduction. 
      White areas are liquid domains and gray areas particles. 
      Holes inside the particles are due to manufacturing and are removed by image processing. 
      \textit{Right}: After image processing pixels belonging to spheres are marked blue and yellow areas
        correspond to liquid clusters.
}
    \label{fig:tomo_slice}
\end{figure}

Everyone who has ever built a sand castle at the beach is familiar with the different mechanical 
properties of wet and dry granular media. Adding a small amount of a wetting liquid to the particles 
allows the formation of capillary bridges between particles 
\cite{Hornbaker_1997,Halsey_1998,Mason_1999,kohonen:04,butt:09,gogelein:10,Herminghaus_2013}. 
These bridges add tensile forces
to the packing, resulting in a significantly increased mechanical stability 
\cite{mason:68,pierrat:97,tegzes:02,Scheel_2004,Fournier:05,Huang:05,richefeu:06,xu:07,moller:07,huang:09,rahbari:09,fiscina:10,Liu_2011,vandewalle:12,fiscina:12,2014_Fall_SlidingFrictionOnWetAndDrySand,bossler:16,kovalcinova:18,liefferink:18}: 
it is simple to build vertical sand castle
walls from wet sand while the same material in its dry state can not form piles with steeper 
slopes than its angle of repose around 30$^\circ$, which is dependent on the friction coefficient of the sand.

However, this intuitive notion of additional tensile forces from capillary bridges has not yet been translated 
into a quantitative theory capable of predicting the properties of a specific wet granular material.
Part of the problem is that while X-ray tomography studies \cite{Scheel_2008,scheel:09,mani:15,saingier:17}
have provided a more detailed picture of the liquid 
morphologies inside wet packings, it is still unclear how much these additional tensile forces also lead to 
geometrical rearrangements of the particles by drawing close-by particles together. 
Such rearrangements can change the mechanical properties of the sample:
The stability of a granular sample is conferred by its force chains
\cite{Majmudar_2005,Puckett_2013,Brodu_15,Daniels_2017,kollmer:18}, 
which consist of lines of contacts. Therefore a change of fabric could modify 
the mechanical behavior independently of the tensile nature of the bridges.

In this work we demonstrate that while adding liquid to a sphere packing does introduce tensile forces, 
it does not significantly change the geometrical arrangement of our sphere packings.
And due to their disability to interlock when compared to other particle shapes, spheres can be expected to be most susceptible to such  geometrical changes.

\section{Experiments}
\subsection{Packing preparation}
We prepare wet and dry packings of approximately 5000 monodisperse Polyoxymethylene (POM) spheres (diameter $d$ = 3.5 $\pm$ 0.022 mm) 
in a plexiglass cylinder of 84 mm diameter  and 150 mm height. In order to avoid crystallization \cite{Hanifpour_2015}
in the dry experiments, the container walls are lined  with bubble wrap foil. Wet packings are prepared by 
distributing Bromodecane between the spheres by continuously rotating and shaking the container.
Bromodecane wets POM surfaces well and provides good X-ray contrast to POM.
We will show below that all particles in contact have liquid bridges. We note our earlier failed attempts 
to use glass particles with a solution of CsCl in water, which did not result in a homogeneous distribution of the liquid.
  
Measurements are performed for liquid volume fractions $\ltotperc=\SI{0}{\%}$, $\ltotperc=\SI{2.1}{\%}$ and $\ltotperc=\SI{3.1}{\%}$,
where $\ltotperc$ is defined as the total liquid volume divided by the total sample volume \cite{Scheel_2004,Scheel_2008}.

In choosing our experimental parameters we need to consider the ratio of gravitational and surface tension forces 
in our sample, which is measured by the E\"otv\"os number
$Eo = \Delta \rho \, g \, d^2 / \sigma$ where $g$ is the gravitational acceleration. 
Inserting our values for the density difference $\Delta \rho$ = 1.4 g/cm$^3$ and surface tension $\sigma$ = 30 mN/m 
we obtain $Eo$ = 5.7, i.e. gravitational forces dominate, but surface tension still plays an important roll. 


The contribution of surface tension to the mechanical stability of the packings becomes
evident when comparing the range of accessible global volume fractions $\phi_g$: 
Wet packings have a volume fractions after preparation in the range 0.575 to 0.591\footnote{It was not feasible to increase $\phi_g$ in wet packings above 0.605 because longer shaking lead to crystallization of the sample and we could not use rough container boundaries as this would have lead to inhomogeneous liquid distributions.}.
Continuous vertical shaking (30 to  6000 seconds, 30 Hz, maximal acceleration 8g) 
increases their $\phi_g$ to only 0.588 - 0.605. In contrast,we could not prepare dry packings 
at $\phi_g$ below 0.61; vertical tapping (10 to 160000 sinusoidal taps with a maximal 
acceleration of 2 $g$) increases  $\phi_g$ up to 0.64 \cite{knight:95,Ribiere_2005} .
These non-overlapping $\phi_g$ ranges of wet and dry packings are a clear testimonial to the influence 
of liquid bridges on the mechanical properties, even at a E\"otv\"os number larger one.

\subsection{Data acquisition}
The internal structure of the packings is evaluated using X-ray tomography; 
wet samples are measured with a CT-Rex (Fraunhofer EZRT,  voxel resolution \SI{35}{\micro\meter}), 
dry samples with a Nanotom (GE Sensing and Inspection, voxel resolution  \SI{64}{\micro\meter}).
The image size after cropping the boundary is $1300 \times 1300 \times 1000$ voxels in both setups.  All wet packings were prepared at least 30 
minutes prior to the measurements in order to allow the bromodecane to equilibrate inside the packing \cite{Scheel_2008,scheel:09}.

Particle center positions are detected for all particles, but the further analysis is limited 
to the $N\approx 800$ particles which are at least 20 mm away from the boundaries, 
using the methods described in \cite{Weis_2017_xray}\footnote{Parameters for image processing: Bilateral filter $\sigma_g = \num{4}$ in units of voxels, $\sigma_p = \num{2700}$ in units of greyvalues and an erosion depth $\lambda=\num{5}$ voxels.}.
Voxels representing bromodecane are detected by a second binarisation with another, higher threshold. 
Figure \ref{fig:tomo_slice} shows cross-sections of the raw and segmented tomographies.
Rendered visualizations of the wet packings can be found at \cite{video1,video2}. 

\subsection{Computing global volume fractions}
The global volume fraction inside our analysis area is the harmonic mean \cite{Aste_2006_volumeFluctuations,Weis_2017_xray} 
of the local volume fractions $\phi_l^i$ of the individual particles $i$ inside that area.
$\phi_l^i$ is calculated as the ratio of the volume  of particle $i$ and the volume
of its Voronoi cell \cite{Voronoi_1909,Rycroft_2009,Weis_2017_pomelo,pomelo_theo1} i.e.~the volume which is closer to this spheres
center than to any other sphere center

\subsection{Computing contact and bridge numbers}
\begin{figure}[t!]
  \centering
   \def\svgwidth{0.45\textwidth}
   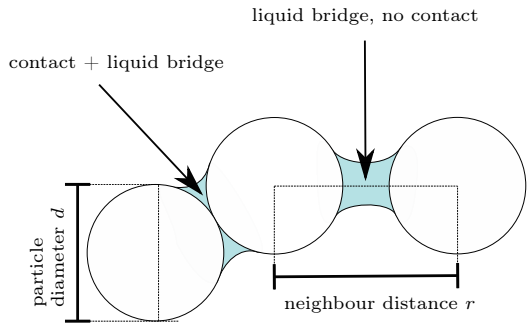
   \caption{ Sketch of the two different types of liquid bridges. In all our experiments, we have found not a single particle contact 
    without an associated liquid bridge.
}
    \label{fig:sketch}
\end{figure}

The next step is the determination of the average contact numbers $\langle Z \rangle$
and the average bridge number $\langle B \rangle$, cf.~figure \ref{fig:sketch}. 
The former describes how many other sphere surfaces a typical particle touches,
the latter counts the number of liquid bridges formed by an average particle. 
$\langle Z \rangle$ is determined using the 'contact number scaling function' (CNS) method 
\cite{Rintoul_1998,Aste_2005a,Aste_2006, Schaller_2013_tomo,Schaller_2015,Weis_2017_xray}; 
this method uses the ensemble of all interparticle distances to determine the correct threshold
distance up to which two particles are considered to be in contact. 
$\langle B \rangle$ is measured by detecting all liquid domains which connect the surface of two spheres.
As expected for a good wetting fluid, for all particle pairs in contact there is also a bridge present at that contact.
We also find that within the central region of the packing less than 1 \%  of all liquid domains touch only one particle; and not a single
liquid domain involves three touching particles, so called trimers. 
The latter is expectd to change for higher values of $\ltotperc$. 

\subsection{Analyzing the shape of the Voronoi cells}
Voronoi cell shape analysis is based on a morphometric Minkowski tensor isotropy analysis \cite{Gerd_2011,Gerd_2013}. 
Specifically, we calculate the volume moment tensors $\mathbf{W_0^{2,0}}(K) := \int_K \mathbf{r} \otimes \mathbf{r} \; \mathbf{dV}$ of each individual Voronoi cell $K$ where the origin is chosen to be the particle center point, $\mathbf{r}$ is the position vector, and $\otimes$ is the symmetric tensor product. 
The shape of the Voronoi cell is then characterized by the ratio of the smallest to largest eigenvalue $\isoidx = \frac{\epsilon_\textnormal{min}}{\epsilon_\textnormal{max}}$ where $\epsilon_\textnormal{min}$, $\epsilon_\textnormal{max}$ are the smallest and largest eigenvalue, respectively; note that the eigenvalues are positive. When applied to the Voronoi partition, these Minkowski structure metrics have become commonly used structure metrics \cite{Gerd:2010,Kapfer_2012,Schaller_2015_nonuniversal}, complementary to other metrics such as the two-point correlation function.

\begin{figure}[b]
    \centering
    \includegraphics{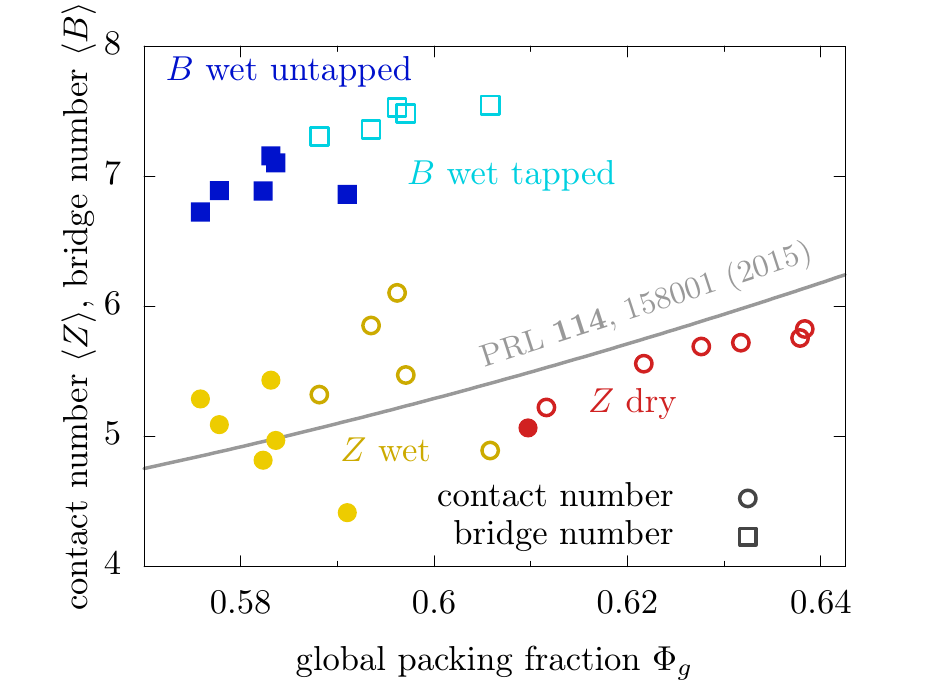}
    \caption{Both bridge number $\langle B\rangle$ and contact number $\langle Z\rangle$ increase with increasing volume fraction.
    Contact numbers are shown as red (dry) and yellow (wet) circles, bridge numbers as blue and turquois squares. 
        Wet measurements correspond to $\ltotperc=\SI{2.1}{\%}$. 
        Open and closed symbols correspond to tapped and untapped system preparation, respectively. 
        }
    \label{fig:bncnphig}
\end{figure}

\section{Comparing wet and dry packings}
\subsection{Contact and bridge number}

Figure \ref{fig:bncnphig} summarizes our results for $\langle Z\rangle$  and  $\langle B\rangle$. 
Our main finding here is that, within experimental noise, the average  $\langle Z\rangle$ of both wet and dry packings seem to be consistent with the previously published empirical model for dry spheres \cite{Schaller_2015}. 
The small but systematic deviations of the dry packing data from the model can be explained 
by the larger asphericity of the 3D printed particles used in reference \cite{Schaller_2015}.
For more details see the supplemental material.

It is also noteworthy that the $\langle Z\rangle$ values of wet packings fluctuate more strongly from preparation to preparation 
than those of dry packings, in contrast to the $\langle B\rangle$ values which display a smoother increase with $\phi_g$. 
Dry packings are hyperstatic, i.e.~the contact forces provide more constraints than what is needed to fix
all the degrees of freedom  of the particles  \cite{schroeter:17}.
In absense of a dominating mechanical constraint, 
the dependence of  $\langle Z\rangle$  on $\phi_g$ originates from geometrical effects
such as volume exclusion \cite{song:08,baule:13}.
In wet packings, hyperstaticity is even stronger than in dry packings due to the addition of the capillary bridges as another force transmission mechanism. 
However, the stronger fluctuations of 
$\langle Z\rangle$ compared to $\langle B\rangle$ indicates that the physics of wet packings
 might be less determined by volume exclusion (which is strongly connected to $\phi_g$) than by 
the preparation dependent interplay between tensile and compressive forces.
 
Figure \ref{fig:bncnphig} shows also that in wet packings each particle has on average two liquid bridges that do not correspond to a particle contact: $\langle B\rangle -  \langle Z\rangle \approx 2$
(keeping in mind that each contact also corresponds to a liquid bridge).
Put differently, the number of force transmission channels for compressive forces (at the contact points) is by two lower than the number of force transmission points for tensile, cohesive forces (all liquid bridges).

\begin{figure}[tb!]
    \centering
    \includegraphics{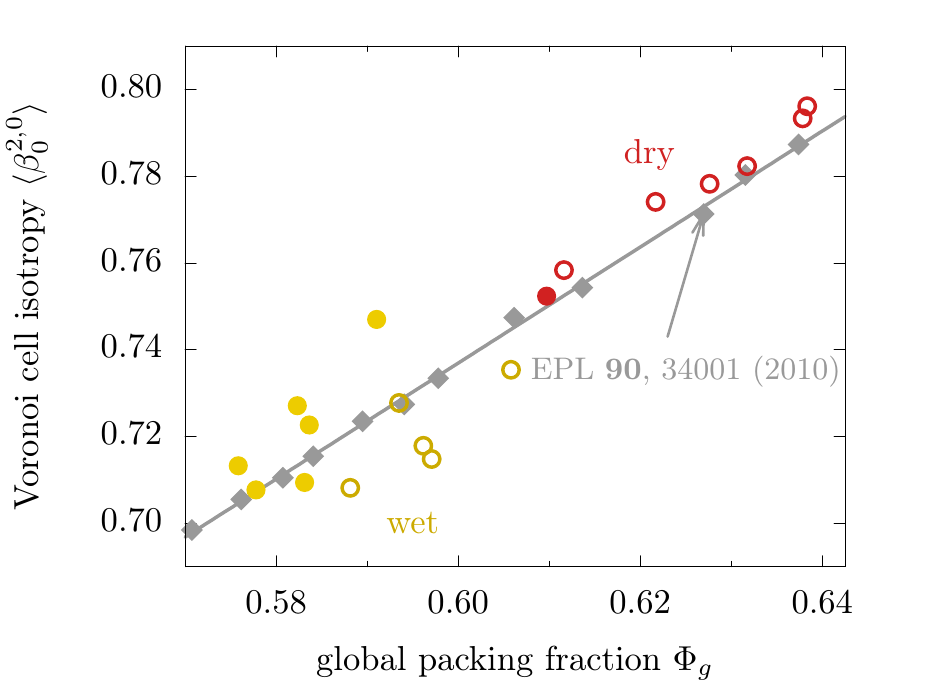}
    \caption{Adding liquid to a packing does not change the shape of the Voronoi cells: the average 
    cell isotropy $\langle\isoidx\rangle$ of dry (red circles) and wet (yellow circles) packings agrees with the previously
     published data (grey diamonds) for dry sphere packings \cite{Gerd:2010}. Open and closed symbols correspond to Fig.\ \ref{fig:bncnphig}.
    }
    \label{fig:packinganisotropy}
\end{figure}

\subsection{Voronoi cell shape}

The Voronoi cell shape analysis shown in figure \ref{fig:packinganisotropy}
reinforces the conclusion that wet packings can be thought of as dry packings with additional liquid bridges introduced between particles that are not in contact but very close by.  
It shows the average isotropy index over all particles $\langle\isoidx\rangle$, calculated for all dry and wet packings. The average packing isotropy $\langle\isoidx\rangle$ allows for no identification of a structural difference between dry and wet packings. Both wet and dry systems coincide with earlier independent results \cite{Gerd:2010}. For the system and length scale studied here, this supports the conclusion that, structurally, the wet packings are 'just dry packings with added liquid bridges', with the presence of the liquid bridges not being accompanied by a significant change in structure of the packing.

Another commonly used metric to study the structure of amorphous packings is the pair correlation function. We show in the appendix that there are again no differences between wet and dry packings
within the experimental resolution of our data.

\begin{figure}[t]
  \centering
  \includegraphics{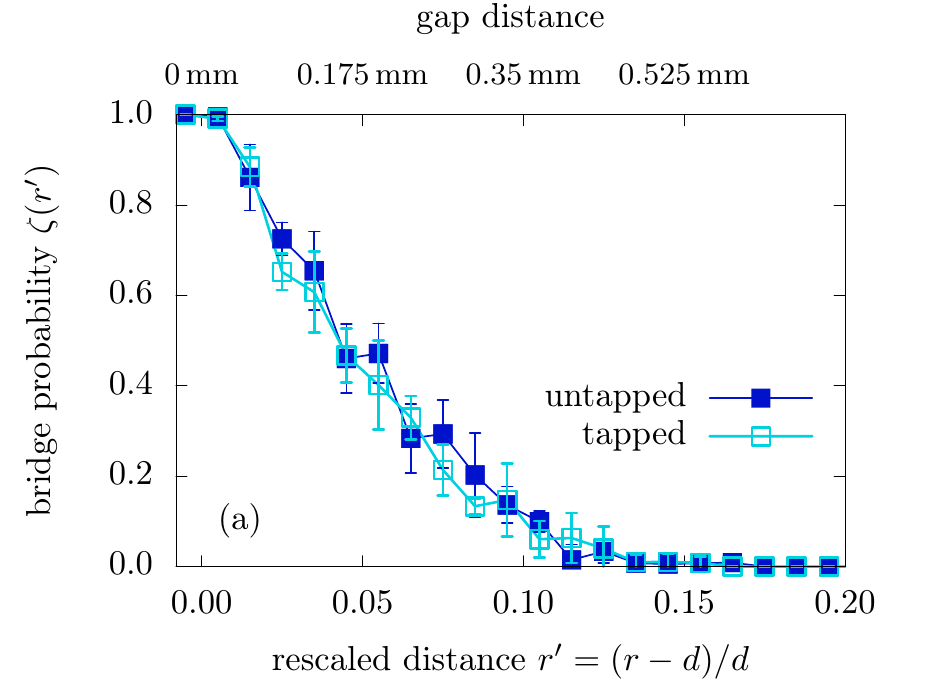}
  \caption{The probability $\zeta$ that a liquid bridge exists between two particles depends only on their
  distance  $r$, not the global volume fraction $\phi_g$ of the packing. Here the distance between spheres is rescaled with the 
  particle diameter $d$: $r'=(r-d)/d$. Typical liquid bridges between particles extend up to 0.15$d$.
  To improve statistics, we have averaged  all tapped  and all untapped
  experiments. The inset provides the proof that this is permissible because  $\zeta(r')$ does not depend on
  $\phi_g$: the slope of the curves, determined from a linear fit to $\zeta(r')$ in the range between 0 and 0.1,
  is constant for all experiments. 
  }
  \label{fig:wetcontactdistances_multiplot_a}
\end{figure}

\section{Probability of a liquid bridge forming between spheres}

While both $\langle Z\rangle$ and $\langle B\rangle$ depend on $\phi_g$, the fraction $\zeta(r)$ of particle pairs 
at distance $r$ that are connected through a liquid bridge (called \textit{bridge probability}) does not. 
This is shown in  figure \ref{fig:wetcontactdistances_multiplot_a} using a rescaled distance $r'$ 
between particles: $r'=(r-d)/d$.  Particles in contact always have a liquid bridge, $\zeta(0)=1$,
as expected for a good wetting liquid. 
Liquid bridges between particle pairs with gap distances $r-d>0.15d$ do  not occur within our packings.
Between these two limits, $\zeta(r)$ decays monotonously with a slope that is independent of $\phi_g$. 

This result supports an inversion of our main argument: while the geometry of the packing does change (increasing
 $\phi_g$ does change the pair correlation function), the probability of formation of liquid bridges seems to be
 unaltered (which points to liquid properties such as surface tension as the main control parameter).

\begin{figure}[t]
    \centering
    \includegraphics{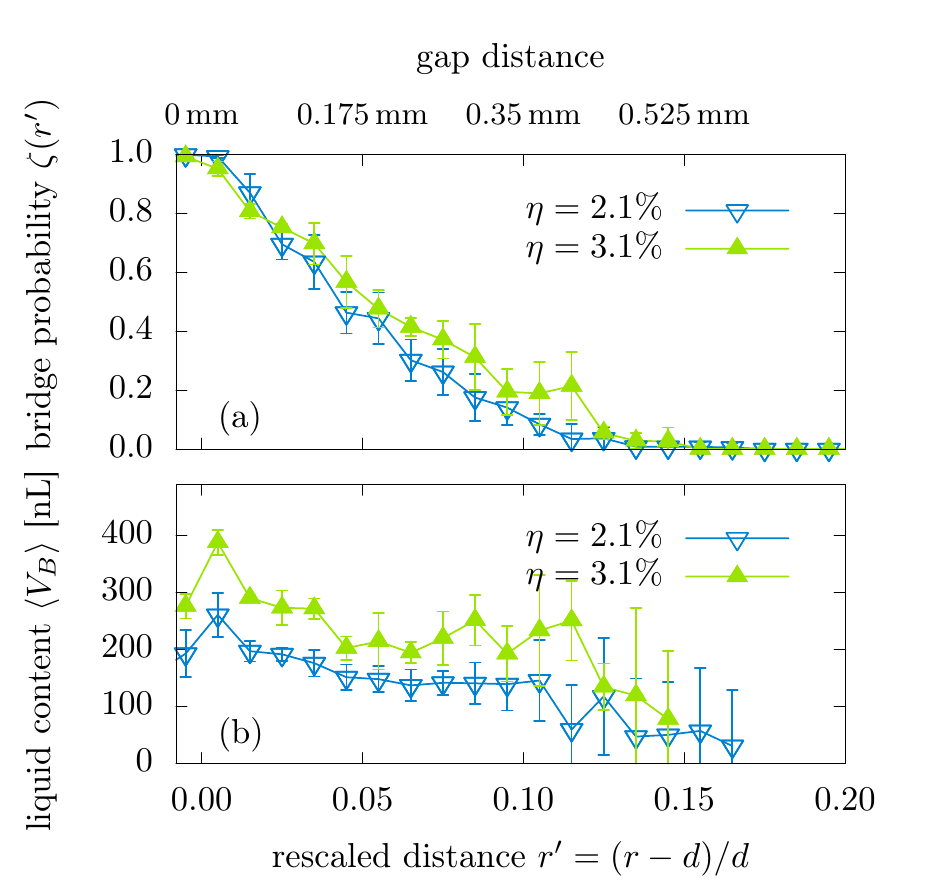}
    \caption{Adding more liquid increases the number of and the average volume inside liquid bridges. But 
    it does not create additional bridges between particles that are further apart. This is shown in panel \textbf{a)}:
    Typical liquid bridges extend up to 0.15 particle diameters regardless if the liquid content
    $\ltotperc$  is $\SI{2.1}{\%}$ or $\SI{3.1}{\%}$.
    \textbf{(b)}: The average volume inside liquid clusters $\langle V_B \rangle (r')$ increases with $\ltotperc$. 
    If no liquid bridges exist in a given distance bin, the corresponding data point is omitted.
    Data is averaged over 11 ($\ltotperc$ = 2.1 \%)  respectively 7 ($\ltotperc$ = 3.1 \%)  experiments. 
    }
    \label{fig:wetcontactdistances_multiplot_b}
\end{figure}

All data for wet packings shown in figures \ref{fig:bncnphig}  to  \ref{fig:wetcontactdistances_multiplot_a} are for a liquid volume fraction $\eta=2.1\%$. 
Figure \ref{fig:wetcontactdistances_multiplot_b} addresses the natural question how the shape and distributions 
of liquid bridges changes when the liquid volume fraction is increased: the main effect is an increase in the volume (and hence shape) 
of the liquid bridges, not the creation of additional longer liquid bridges.

Figure \ref{fig:wetcontactdistances_multiplot_b} (a) shows that while the bridge probability $\zeta(r)$ does increase slightly
for $\eta$ = 3.1 \%, there is no emergence of longer bridges: $\zeta(r)$ still drops to zero for  $r-d\approx 0.016 d$.
The main effect of increasing $\eta$ is a proportional increase in size of the liquid bridges: 
within statistical accuracy the ratio of the
average bridge volumes $\langle V_B\rangle$ at $\eta=2.1\%$ and $\eta=3.1\%$ corresponds to the ratio of added bromodecane 
for all values of $r'$.  
It is a worthwhile future question beyond the scope of this article to investigate in more detail the shape and volume of the liquid bridges (considering the resolution of our CT data, details of the shape can be  probed; a volume of $200 nL$ corresponds to approximately 4000 voxels).

\section{Conclusion}
 We have found that the structure of the wet packings to be very similar 
to that of a dry packing at the same packing fraction. This is a surprising result considering that 
the presence of the wetting liquid has a clear influence on the mechanical properties. 
Our conclusions depend almost certainly on the sphere diameter because the relative importance of surface tension forces 
compared to gravitational forces increase when $d$ decreases. 
Our results are for beads of diameter 3.5 mm where liquid bridges are relevant for the mechanical properties, despite not affecting the structure of the bead pack. Future research should explore the limit of smaller particles, where the presence of a wetting liquid can stabilize packings at substantially lower packing fractions than what can be reached in dry packings. A second important 
avenue for future research will be non-spherical particles which introduce not only new liquid bridge geometries but also 
interesting new geometrical features in dry packings.

\section*{Acknowledgements}
We are grateful to Martin Brinkmann (Saarbr\"ucken) for advice and discussion, and Klaus Mecke and Thorsten P\"oschel (both Erlangen) 
and Stephan Herminghaus (G\"ottingen) for support. 
We acknowledge funding by the German Science Foundation (DFG) under grant SCHR-1148/3-2 within the research group 'Geometry and Physics of Spatial Random Systems' 
an through the Cluster of Excellence 'Engineering of Advanced Materials'.  
GEST acknowledges support through a collaboration scheme of Universities Australia and the German Academic Exchange Service (DAAD). 

\appendix
\section{ Comparing $g(r)$ of dry and wet packings}

\begin{figure}[h]
\includegraphics{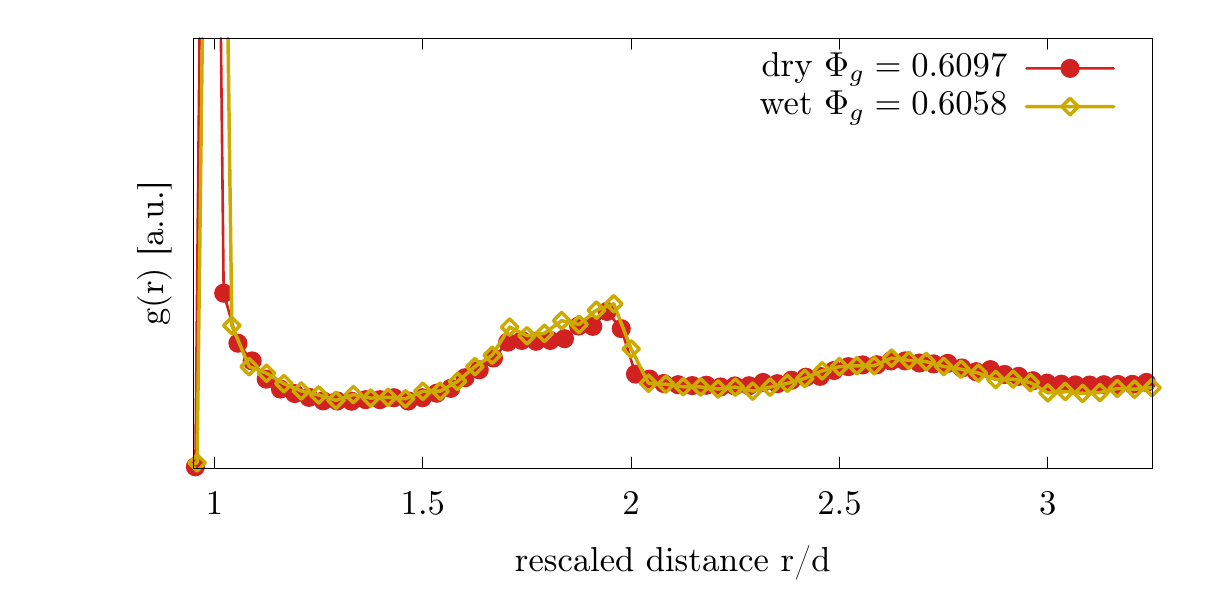}
\caption{Within the statistics of our measurement no structural differences are revealed by $g(r)$.
    Pair correlation function $g(r)$ comparison between a wet packing at $\Phi_g = 0.6058$ and a dry packing at $\Phi_g=0.6097$. 
}
\label{fig:gr}
\end{figure}

The pair correlation function $g(r)$ is a commonly used structural measure for granular packings.
It is known that the features of the pair correlation function $g(r)$ of granular packings depend on the global packing fraction $\Phi_g$ of the packing. 
Packing fractions for wet and dry packings did not overlap in our experiments.
Therefore we show the packings with the smallest difference in global packing fraction, when comparing the pair correlation function $g(r)$ for wet and dry packings in figure \ref{fig:gr}. 
The wet packing has a value of $\Phi_g=0.6058$ and the dry packing of $\Phi_g=0.6097$.
As it can be seen in figure \ref{fig:gr} no structural differences are revealed by $g(r)$

\section{ Comparing the particles in this work and in Schaller \textit{et al.} PRL 2015}

The 3D printed particles used in Schaller \textit{et al.} PRL 2015 \cite{Schaller_2015} are characterized by an average asphericity $\beta_0^{2,0}$ (The asphericity is evaluated on the particle shape, not its Voronoi cell) of $0.96 \pm 0.01$ whereas the particles used here are more spherical with $\beta_0^{2,0}\approx 0.992 \pm 0.004$.)
It is known that with increasing particle asphericity (decreasing particle aspect ratio) the average contact number $\langle Z \rangle$ increases. 
This explains the deviations in the contact number between the particles used in this article and in Schaller \textit{et al.} PRL 2015.

\begin{figure}[h]
\includegraphics{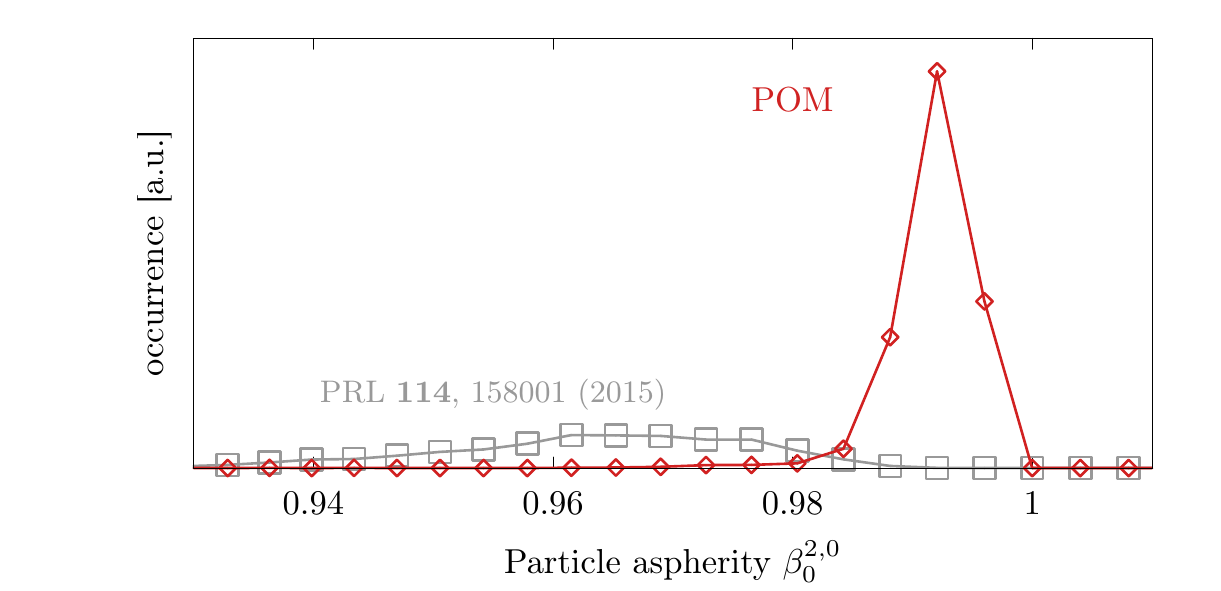}
\caption{
    The particle asphericity distribution for the particles used in this work (POM) is significantly closer to one than for the 3D printed particles used in \cite{Schaller_2015}.}
\label{fig:spheres_non_perfect}
\end{figure}

\end{document}